\documentclass[prd,superscriptaddress,onecolumn,showpacs,11pt,msmath,
preprintnumbers,showkeys]{revtex4}
\usepackage{wrapfig,rotating}
\usepackage{graphicx,epsfig,color}

\makeatletter

\newenvironment{figurehere}
{\def\@captype{figure}}
{}
\makeatother

\usepackage{graphicx}
\usepackage{dcolumn}
\usepackage{bm}

\def\beq{\begin{equation}}
\def\eeq{\end{equation}}
\def\beeq{\begin{eqnarray}}
\def\eeeq{\end{eqnarray}}

\def\cO#1{{\cal{O}}\left(#1\right)}

\def\2GPD{$_2\mbox{GPD}$}

\def\12{$1\otimes 2$}
\def\22{$2 \otimes 2$}

\def\Qsep{Q_{\mbox{\rm\scriptsize sep}}}
\def\Qsep2{Q^2_{\mbox{\rm\scriptsize sep}}}

\begin{document}

\title{Double parton interactions in $\gamma p, \gamma A$ collisions in the direct photon kinematics.}
\pacs{12.38.-t, 13.85.-t, 13.85.Dz, 14.80.Bn}
\keywords{pQCD, jets, multiparton interactions (MPI), LHC, TEVATRON}

\author{B.\ Blok$^{1}$,
 M.\ Strikman$^{2}$
\\[2mm] \normalsize $^1$ Department of Physics, Technion -- Israel Institute of Technology,
Haifa, Israel\\
\normalsize $^2$ Physics Department, Penn State University, University Park, PA, USA}

\begin{abstract}
We derive expressions for the   differential distributions and the total cross section of double- parton interaction in direct photon interaction with proton and nuclei. We demonstrate that  in this case the cross section is more directly related to the nucleon generalized parton distribution than in the case of double parton interactions in the proton - proton collisions.
We focus on the production of two dijets each containing charm (anticharm) quarks and carrying $x_1,x_2>0.2$
fractions of the photon momentum. Numerical results are presented for the case of  $\gamma p$ collisions at
LHeC, HERA and in the
 ultraperipheral $AA$ and $pA$
collisions at the LHC.
We find that the events of this kind would be abundantly produced at the  LHeC. For  $\sqrt{s}=1.3$ TeV the expected rate is $2\cdot 10^8$ events for the  luminosity $10^{34}$ cm$^{-2}s^{-1}$, the running time of $10^6$ s and  the  transverse cutoff of $p_t>5$ GeV. This would make  it feasible to use these processes for the model independent determination of two parton GPDs in nucleon and in nuclei.
For HERA the total accumulated number of the events is also high, but efficiency of the detection of charm seems too low to study the process. We also find   that a significant number of such double parton interactions should be produced in $p - Pb$ and $Pb- Pb$ collisions at the LHC: $\sim 6\cdot 10^4$ for $Pb-Pb$, and $\sim  7 \cdot 10^3$ for $p-Pb$  collisions for the same transverse momentum cutoff. \end{abstract}

  \maketitle
\thispagestyle{empty}

\vfill

\section{Introduction}

{\em Multiple hard parton interactions}\/ (MPI) started to play an important role in the description of the inelastic $pp$ collisions at the collider energies.
 Hence, although the studies of MPI began in eighties \cite{TreleaniPaver82,TreleaniPaver85,mufti,dDGLAP}, they attracted a lot of theoretical and experimental attention only recently.
Extensive theoretical studies were carried out in the last decade, both for pp collisions
\cite{Treleani,Diehl,DiehlSchafer,Diehl2,Wiedemann,Frankfurt,Frankfurt1,SST,stirling,stirling1,Ryskin,Berger,BDFS1,BDFS2,BDFS3,BDFS4,Gauntnew,Gauntadd,Sjodmok}, and for $pA$ collisions \cite{ST,BSW,S}.
Attempts have been made to incorporate multi-parton collisions in the Monte Carlo  event generators \cite{Pythia,Herwig,Lund,BFS}.

MPIs can serve as a probe for {\em non-perturbative correlations}\/ between partons in the nucleon
wave function and are crucial for determining the structure of the underlying event at the LHC energies.
They constitute an important background for the new physics searches at the LHC.
A number of experimental studies were performed at the Tevatron \cite{Tevatron1,Tevatron2,Tevatron3}.
New
experimental studies are underway at the LHC \cite{Atlas,cms1,cms2,cms3}.

\par
The analysis of the experimental data indicates \cite{Frankfurt,BDFS1}  that the rate of such collisions exceeds significantly a naive expectation based on the picture of the binary collisions of the uncorrelated partons of the  nucleons (provided one uses information from HERA on the transverse distribution of gluons in nucleons).

\par In the parton model inspired picture MPI occur via collisions of the pairs of partons: the \22  mechanism (collision of two pairs of partons).  In pQCD the picture is more complicated  since the QCD evolution generates short-range correlations between the partons (splitting of one parton into two,...) --the \12  mechanism\cite{BDFS2,BDFS3,BDFS4}.  It was demonstrated that account of these pQCD correlations enhances the rate of MPI as compared to the parton model by a factor of up to two and may explain discrepancy of the data \cite{Tevatron1,Tevatron2,Tevatron3,Atlas,cms1,cms2,cms3} with the parton model. (A much larger enhancement recently reported in the double $J/\psi$ production \cite{DO14} can hardly be explained by this mechanism).

Presence of two mechanisms and limited knowledge of the nucleon multiparton structure makes a unique interpretation of the data rather difficult.

Hence here we propose to study the MPI process of
$\gamma p (A)$ interaction  with production of four jets in the kinematics where two jets carry most of the light cone fraction of the photon four momentum - direct photon mechanism. In this process the \22 mechanism is absent and the only process which contributes is
 an analog of the \12 process.Since in the proposed kinematics the contribution of the resolved photon is strongly suppressed the cross section in  the leading log approximation (LLA)  i.e. summing leading collinear singularities is expressed through the integral over  two particle GPDs,
$_2D(x_1,x_2,Q_1^2,Q_2^2,\Delta)$, introduced in \cite{BDFS1}. This is in difference from the case of $pp,pA$ scattering  where \22  contribution  is proportional to  a more complicated  integral with the integrant proportional to the  product of two double
parton GPDs.

\par The main goal of the present paper  is to show that processes with direct photon in photon proton  collisions provide  a golden opportunity for the
model independent determination of the double parton distributions $_2D$, free of the ambiguities inherent in $pp/pA$ scattering \cite{BDFS3}.
We will  consider the
 process of the interaction of the
real/quasireal photon  with proton with production of
  two pairs of hard jets in the  back to back kinematics with each dijet consisting of a heavy (charm) quark and gluon jets  (see Figs. 1 and 2).
  We focus on the production of charm to suppress the contribution of the resolved photons.

   \begin{center}
 \begin{figurehere}
   \includegraphics[height=5.6cm]{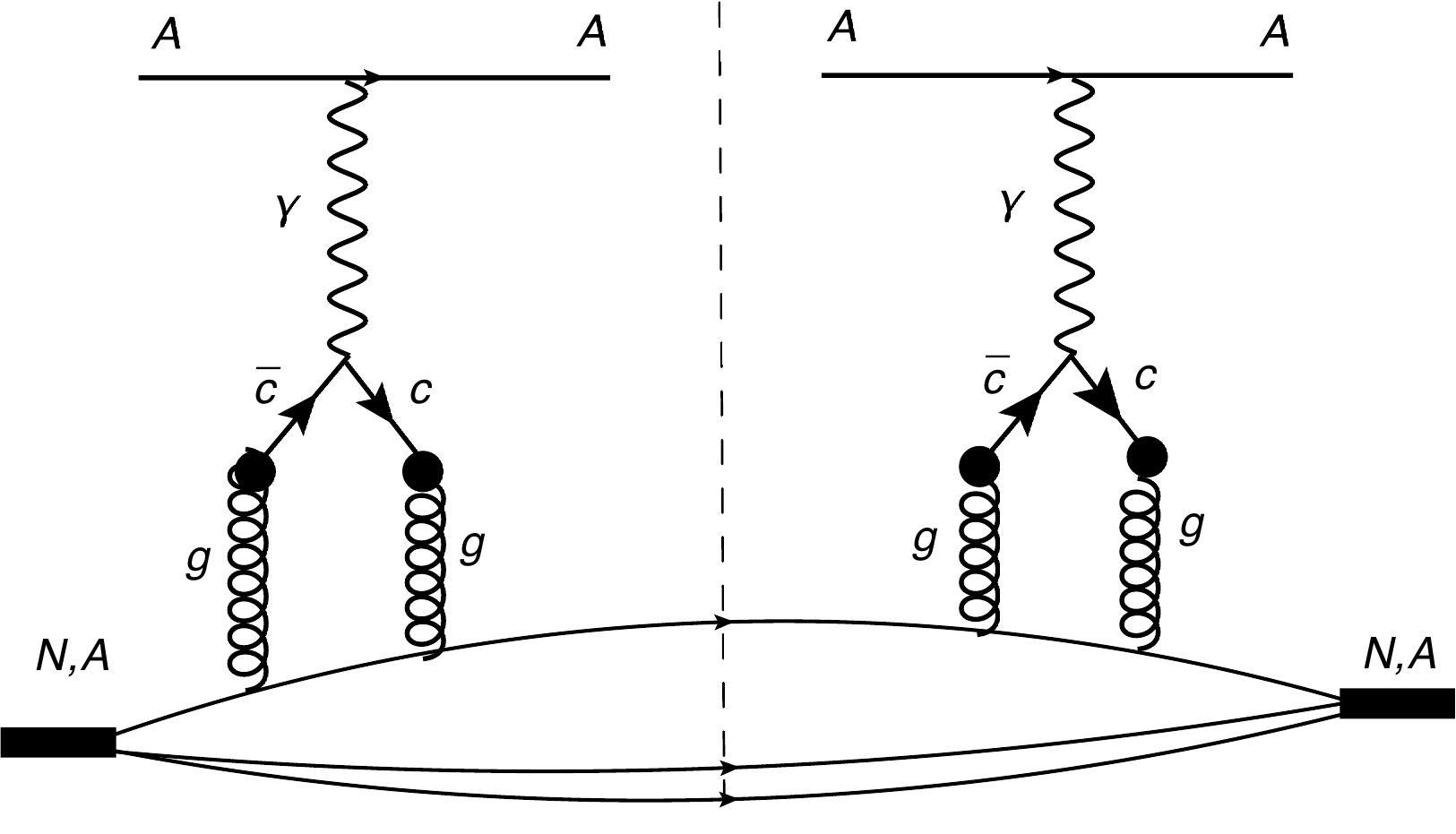}
   \caption{  \label{kin}Fig. 1-MPI two dijet photoproduction -$Ap$}
 \end{figurehere}
 \end{center}
\begin{center}
\begin{figurehere}
 \includegraphics[height=5.5cm]{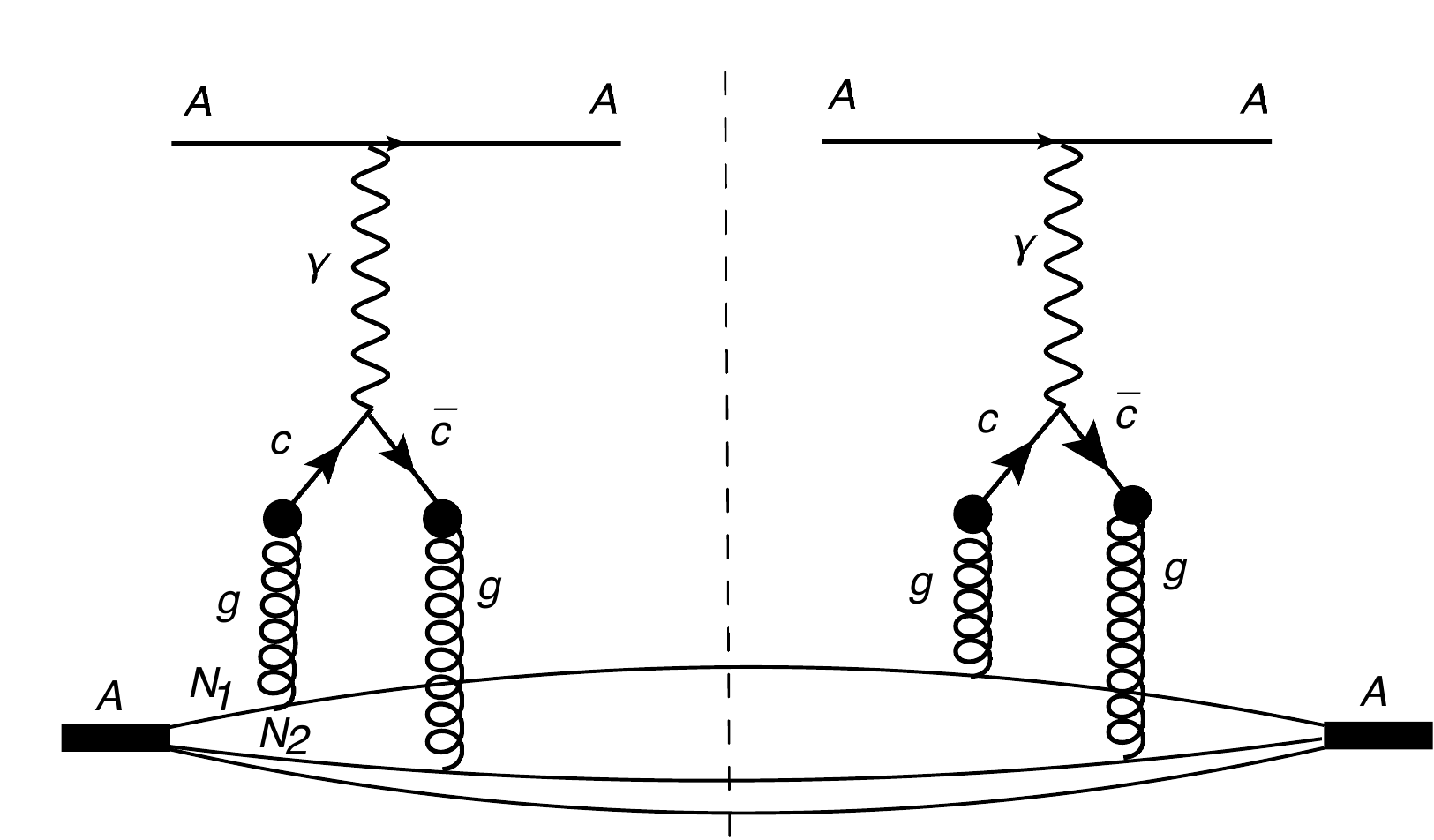}
\caption{\label{Fig2} MPI two dijet photoproduction-$AA$ }
\end{figurehere}
\end{center}
In the discussed process a  $c\bar c$ pair is produced in the photon fragmentation region, while two gluon jets are created predominantly  in the target region,
so that there is a  large rapidity gap between the
gluon and quark jets. The gluon and c-quark jets are approximately balanced pair vice. The cross section of the analogous process in $pp$ collisions is influenced by parton correlations in both nucleons participating in the process,
while in the case of the photon the cross section depends only on
the integral over one wave function.
 The reason is  that the process
involves only one GPD from the nucleon, while the upper part of the diagram 1 is determined by the hard physics of the photon splitting to $Q\bar Q$ pair in an unambiguous way.
It does not  involve the scale $Q_0^2$ that separates perturbative and nonperturbative correlations in a nucleon. Thus the cross section of such a process is directly expressed through the
nucleon
 double GPD. 
 Hence the measurement of the discussed cross section would allow
 to
 perform a nearly model independent analysis of
 DPI in  $pp$ scattering.
We will
 demonstrate
 below that it would be possible also to study these processes at the future electron - proton / nucleus colliders. It maybe possible also to investigate these processes in
$AA$ and $pA$ ultraperipheral collisions at the LHC.

\par Here we will
 consider the MPI rates for all three types of processes mentioned above, $\gamma p, AA, pA$. We will restrict ourselves to the   kinematics $x_1,x_2>0.2$, thus guaranteeing
the dominance of the direct photon contribution (For  this cutoff the direct photons contribute 60\% of the
dijet
 cross section).
For a lower $x_i$ cutoffs the relative contribution of the direct photon mechanism  rapidly decreases for transverse momenta under consideration.

We will demonstrate that for the LHeC collider  energies
$\sqrt{s}=1300$ GeV the rate of the discussed reaction will be very high:  $2\cdot 10^8$ events per $10^6$ s for  the luminosity $10^{34}$ cm$^{-2}$s$^{-1}$ and   $p_t>5$ GeV. The relative rate of MPI to
 to dijets  is found to  be  $0.045\%$. In principle a large number of events in the discussed kinematics was  produced at HERA: $\sim 1.2\cdot 10^5$
 for the total luminosity $1$ fb$^{-1}$.  However the efficiency of the detection of  $D^*$  was pretty low \cite{Zeus} so it appears that it would be very difficult to 
 study  the discussed process using 
 the HERA data.

Another way to observe the discussed process in the near future maybe possible - study of MPI in the  ultraperipheral $pA, AA$ processes at the full LHC energy.
For example, for $p_t>5$ GeV, we have $\sim 6\cdot 10^4$ events for $AA$, and $\sim 6.6\cdot 10^3$ events  for $pA$ scattering where we used luminosities
 $\sim 10^{27}$ ($AA$) , $10^{29}$ cm$^{-2}$s$^{-1}$ ($pA$) and running time of $10^6$ s.
In the discussed kinematics  MPI events constitute  $\sim 0.04\%$ ($\sim 0.02\%$, $\sim 0.0125\%$) of the dijet events for $AA , pA$ collisions respectively
 for the same jet cutoff. These fractions  decrease rather rapidly with $p_t$ increase.
\par Of course, the MPI processes are contaminated by the leading twist 4 jet production the so called
 2 to 4 processes. However it is possible to argue that in the  back
to back kinematics the contribution of these processes(see Fig.3)  are parametrically small in a wide region of the  phase space \cite{BDFS2}. (Moreover for the $AA$ collisions there is
an  additional combinatorial $A^{1/3}$ enhancement over parasitical 2 to 4 contributions \cite{ST,BSW}. )
Indeed, a detailed MC simulation analysis was done
using Pythia and Madgraph for $pp$ collisions \cite{Tevatron1,Tevatron2,Tevatron3}.
These authors have demonstrated that it is possible to introduce observables that are dominated by MPI in the back to back kinematics, thus allowing to
measure MPI cross sections, as distinct from 2 to 4 processes. Of
 course, further work is required in this direction, especially including NLO effects.
\begin{center}
\begin{figurehere}
 \includegraphics[height=5.5cm]{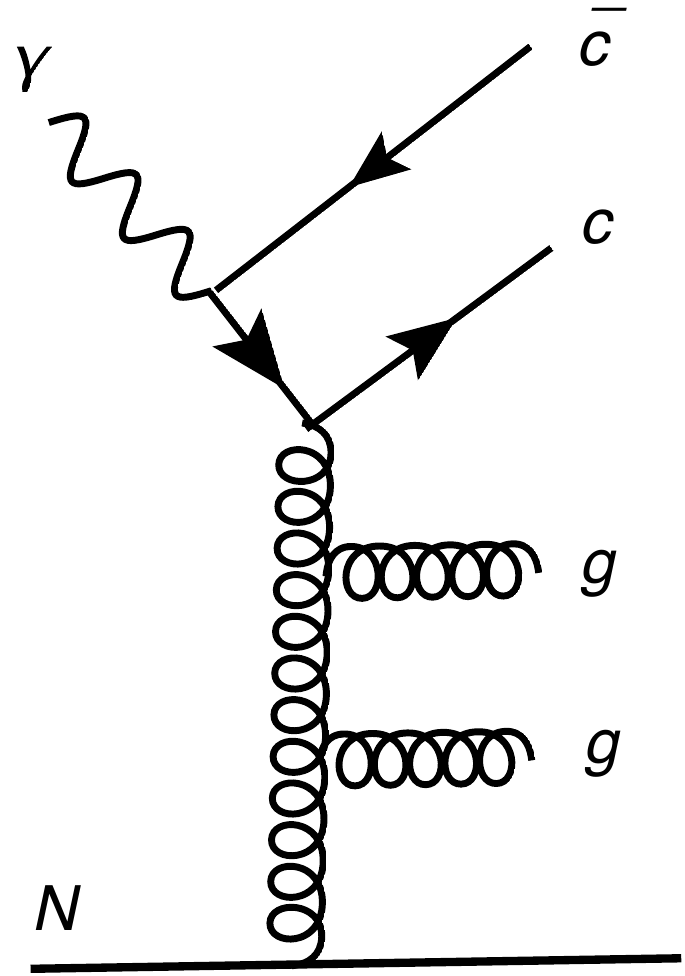}
\caption{\label{Fig4}The
\22 processes-competing kinematics for two  dijets.}
\end{figurehere}
\end{center}

\par Note that MPI in the photon proton collisions were also studied in \cite{BFS}. These authors considered resolved photon kinematics, which is very different,
from the one that is considered here. So there is no overlap with the present study.

\par The paper is organized as following.
In section 2 we calculate the MPI contribution to
$\gamma+p\rightarrow c+\bar c+g_1+g_2+X$ process in the back to back kinematics. In section 3 we calculate the rates of the discussed process for
$ep$ collisions at LHeC and HERA, and for ultraperipheral  $pA$ and $AA$ collisions at the LHC. In section 4 we carry the numerical simulations for realistic parameters corresponding to LHC and HERA runs. The results are summarized
in section 5.

\section{Basic formulae for MPI in the direct photon - proton scattering.}

\subsection{Parton Model.}
\par First we consider the process of production of two dijets with single charm in each pair (Fig.~4a) in the parton model.
In this case the process is essentially the same as the one
 already considered in ref. \cite{BDFS2}. The only
  difference is  that the parton created in the
split vertex is a charmed quark
-- antiquark pair.  The corresponding kinematics is depicted in Fig.~4a, and is analogous to the  \12 transition in $pp$ interactions.
Let us parameterize the momenta of quarks and gluons using Sudakov variables ($k_1,k_2$ are momenta of virtual charm quarks and antiquark  of the  $q\bar q $ pair  and $k_3,k_4$ are the gluon momenta).
Let us analyze the lowest order  amplitude shown in Fig.~4a for the  double hard collision which involves parton splitting.
\begin{center}
\begin{figurehere}
 \includegraphics[height=5.5cm]{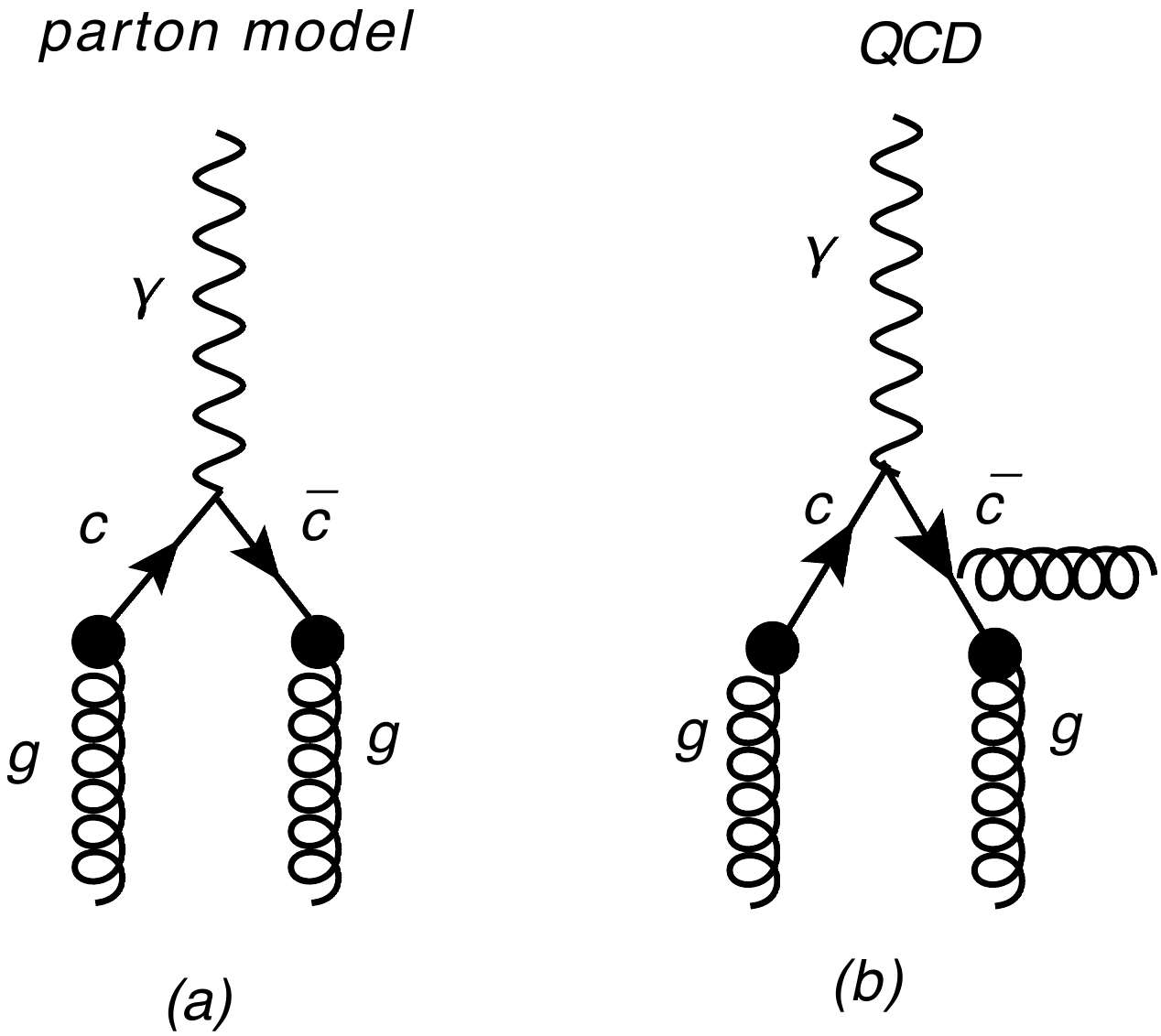}
\caption{\label{Fig3} parton model (a) and pQCD diagrams  (b) for the MPI production of c, $\bar c$ + 2 gluon jets.}
\end{figurehere}
\end{center}

We decompose parton momenta $k_i$ in terms of the so called Sudakov
variables using the light-like vectors $q$ and  $p$ along the incident photon and proton momenta:
\begin{eqnarray}
k_1 &=& x_1q + \beta p + k_{\perp}, \quad  k_3 \simeq (x_3-\beta)p; \nonumber \\
k_2 &=& x_2q - \beta p - k_{\perp}, \quad k_4 \simeq (x_4+\beta)p; \nonumber \\
 \vec{k}_\perp &=& \vec{\delta}_{12} = -\vec{\delta}_{34} \> (\delta'\equiv0); \> k_0 \simeq(x_1+x_2)q. \nonumber
\end{eqnarray}
Here $k_0$ is the  momentum  of the  quasireal photon.
We can neglect the charm quark masses except while dealing with infrared singularities.
The light-cone fractions $x_i$, (i=1,..4), are determined by the jet kinematics (invariant masses and rapidities of the jet pairs).

The fraction $\beta$ that measures the {\em difference}\/ of the  longitudinal momenta of the two partons coming from the hadron, is arbitrary.
The fixed values of the parton momentum fractions $x_3-\beta$ and $x_4+\beta$ correspond to the plane wave description of the scattering process in which the longitudinal distance between the two scatterings is arbitrary. This description does not correspond to the physical picture of the process we are discussing, where two partons originate from the same bound state.
In order to ensure that  partons $3$ and $4$ originate from the  {\em same hadron}\/ of a finite size, we have to introduce
integration over
$\beta$ in the {\em amplitude}, in the  region $\beta=\cO{1}$, as  was explained in detail in \cite{BDFS2}.
\medskip

The Feynman amplitude contains the product of two virtual propagators. The virtualities $k_1^2$ and  $k_2^2$ in the denominators
of the propagators can be written  in terms of the Sudakov variables as
\beq
  k_1^2 = x_1\beta s -k_\perp^2, \>\>  k_2^2 = -x_2\beta s -k_\perp^2, \nonumber
\eeq
where $s=2(p_ap_b)$ and $k_\perp^2\equiv (\vec{k}_{\perp})^2>0$ the square of the two-dimensional transverse momentum vector.

The singular contribution we are looking for originates from the region $\beta\ll1$. Hence
the precise form of the longitudinal smearing
does not play role and the integral over $\beta$ yields the amplitude $A$
\beq
A\sim\!\! \int\! \frac{d\beta}{(x_1\beta s -k_\perp^2-m^2_c +\!i\epsilon)(-x_2\beta s -k_\perp^2- m^2_c+\!i\epsilon)}
=  \frac{2\pi i N}{(x_1\!+\!x_2)}\frac{1}{k_\perp^2+m^2_c}. \nonumber
\eeq
The numerator of the full amplitude is proportional to the {\em first power of the transverse momentum}\/ $k_\perp$.
As a result, the squared amplitude (and thus the differential cross section) acquires the necessary factor $1/\delta^2$
that enhances the back-to-back jet production.
\par The integration over $k_t$ gives a single log contribution to the cross section $\alpha_{\rm em}\log(Q^2/m_c^2)$ where $Q$ is the characteristic transverse scale of the hard  processes.
Note, that strictly speaking the answer is proportional to $\delta(\vec k_{1t}+\vec k_{2t})/(k_{1t}^2+m^2_c)$. The parton model answer is only single collinearly enhanced,
while we are looking for the double collinear enhanced contributions \cite{BDFS2}. It is well known that these contributions originate  from the gluon dressing of the parton model vertex,
with the $\delta$ function becoming a new pole.
\subsection{Accounting for the gluon radiation.}
A typical lowest  order QCD diagram which accounts for the gluon emission
is presented in Fig.~4b. The compensating gluon relaxes the transverse momentum $\delta-$ function. Note however that
the gluon can not be emitted from a photon, while in the $pp$ case such emissions contribute, since the splitting parton carries color. This eliminates the so called short split contribution,
which is present in the case of hadron-hadron scattering. The rest of the calculation is completely analogous to the  "long split" calculation in the $ pp$ case.

Thus using Eqs. 25,26 in \cite{BDFS2} we can write right away
 the differential cross section as

\begin{widetext}
\beeq\label{eq:DD31T}
\pi^2\frac{d\sigma^{(3\to4)}_1}{d^2\delta_{13}\, d^2\delta_{24}} &=&
 \frac {d\sigma_{{\mbox{\scriptsize part}}}
 } {d\hat{t}_1\,d\hat{t}_2}
\cdot \frac{\partial}{\partial\delta_{13}^2}
 \frac{\partial}{\partial\delta_{24}^2} \bigg\{
{}_{[1]}\!D_a^{1,2}(x_1,x_2;\delta_{13}^2, \delta_{24}^2 ) \cdot {}_{[2]}\!D_b^{3,4}(x_3,x_4;\delta_{13}^2, \delta_{24}^2 )
\nonumber\\
&\times&  S_1\left({Q^2},\delta_{13}^2 \right)S_3\left({Q^2},{\delta_{13}^2}\right)
 \cdot S_2\left({Q^2},{\delta_{24}^2}\right) S_4\left({Q^2},\delta_{24}^2\right)   \bigg\} ,
\eeeq
\end{widetext}
where $S_i$ are the quark ($S_1,S_2$) and gluon ($S_3,S_4$) Sudakov form factors \cite{DocBook,DDT}:

\begin{eqnarray}\label{eq:Sudakovs}
S_q(Q^2,\kappa^2) &=& \exp\left\{- \int_{\kappa^2}^{Q^2}\frac{dk^2}{k^2}\frac{\alpha_s(k^2)}{2\pi} \int_0^{1-k/Q} dz\, P_q^q(z) \right\}, \\
S_g(Q^2,\kappa^2) &=& \exp\left\{- \int_{\kappa^2}^{Q^2}\frac{dk^2}{k^2}\frac{\alpha_s(k^2)}{2\pi}
\int_0^{1-k/Q} dz\left[ zP_g^g(z) + n_fP_g^q(z)\right] \right\}.
\end{eqnarray}

\noindent Here $P_i^k(z)$ are the non-regularized one-loop DGLAP splitting functions (without the ``+'' prescription):

\begin{eqnarray}
\label{eq:SPLITS}
  P_q^q(z) & =& C_F \frac{1+z^2}{1-z},  \qquad  \quad\quad \>  P_q^g(z) =   P_q^q(1-z), \nonumber\\[10pt]
  P_g^q(z) & =& T_R\big[ z^2+\! (1\!-\!z)^2\big], \quad\, P_g^g(z) =  C_A\frac{1+\!z^4 +\! (1\!-\!z)^4}{z(1-z)}.\nonumber\\
\end{eqnarray}

The upper limit of the integration over $z$ properly regularizes the soft gluon
singularity, $z\to 1$
(in physical terms, it can be viewed as a condition that the energy of the gluon should be larger than its transverse momentum, \cite{DDT}).
The function $_1D$ now corresponds to the photon split into the  charm anticharm pair.
 Moreover, since we are looking for the
 production of the $c\bar c$ pair in the photon
 fragmentation region, we can
neglect all 
processes except a possible emission of the compensating gluon  by the $c\bar c)$- quark.
Hence we obtain
\beq\label{eq:termPT}
{}_{[1]} \!D(x_1,x_2;   q_1^2,q_2^2; \vec{\Delta}) =  \int_{\Delta^2}^{\min{(q_1^2,q_2^2)}}\frac{dk^2}{k^2+m^2_c}\frac{\alpha_{em}}{4\pi}
 \times  \int\!\frac{dz}{z(1-z)}
R\!\!\left(z\right) \> G^{q}_{q'}\left(\frac{x_1}{z};q_1^2,k^2\right) G^{q}_{q'}\left(\frac{x_2}{(1-z)};q_2^2,k^2\right).
\eeq
The function $R(z)$ is the $q\bar q \gamma$ vertex\cite{peskin}.
\beq
R(z)=z^2+(1-z)^2.
\label{2}
\eeq
The $\Delta$-dependence of ${}_{[1]} D$ is very mild as it emerges solely from the lower limit of the logarithmic transverse momentum integration $Q^2_{\min{}}$.
Here $G^q_q$ is a quark-quark evolution kernel. In the  LLA for hard scale $Q^2\gg m^2_c$ we can use the kernel for massless quarks.

Above we have calculated the differential MPI distributions. We now can integrate the cross section obtaining
\beq
  \frac{d\sigma(x_1,x_2,  x_3,x_4)}{d\hat{t}_1\,d\hat{t}_2}
 = \>\frac{d\sigma^{13}}{d\hat{t}_1}\, \frac{d\sigma^{24}}{d\hat{t}_2}  \int \frac{d^2\Delta}{(2\pi)^2}
{} _{[1]}\!D_{a}(x_1,x_2;Q^2,Q^2)\,  _{[2]}\!D_{b}(x_3,x_4;Q_1^2,Q_2^2;\Delta^2).
\eeq

 Note that we write here the dijet differential cross sections $\frac{d\sigma}{d\hat{t}_1}$ without including  the corresponding PDF factors.

We see that the cross section is unambiguously determined by the integral of $_2$GPD over $\Delta^2$. The factor $_1D$ is given by eq.
\ref{eq:DD31T} (with $\Delta^2=0$) and does not pose any infrared problem, in difference from the  $pp$ case.

\section{Physical kinematics.}
\par There are three possible applications of our formalism--   collisions at HERA and future $ep/eA$ colliders
 and ultraperipheral  $AA$ and $pA$ collisions at LHC.
\subsection{$\Delta$ dependence of input double GPDs.}
\subsubsection{The $\gamma p$ case.}
\par In order to estimate whether it is feasible to observe the  MPI events discussed in the previous section, we have to calculate the double differential
cross section and then to convolute it with the photon flux.

 For the case of the proton target
 we have
\beq
\frac{d\sigma}{dx_1x_2dx_3dx_4dp_{1t}^2dp_{2t}^2}=D(x_1,x_2,p_{1t}^2,p_{2t}^2)G(p_{1t}^2,x_3)G(p_{2t}^2,x_4)\frac{d\sigma}{dt_1}\frac{d\sigma}{dt_2}\int \frac{d^2\Delta}{(2\pi)^2} U(\Delta).
\eeq
Here we carried the integration over the momenta $\Delta$ conjugated to the distance between partons,
 obtaining the last multipliers in the equations above. This integral measures the parton wave function at zero
 transverse separation between the partons   and hence it is sensitive to short-range parton-parton correlations.

  For $\gamma p$ case the factor $U(x_1,x_2,\Delta)$, in the approximation when two gluons are not correlated, is equal to a product of two gluon form factors of the proton:
 \beq
  U(\Delta,x_3,x_4)=F_{2g}(\Delta,x_3) F_{2g}(\Delta,x_4).
  \label{U}
  \eeq

 For the numerical estimates we use  the following approximation for $_2GPD$ of the nucleon:
 \beq
 _2D(x_3,x_4,p_{1t}^2,p_{2t}^2,\Delta)=G(x_3,p_{1t})G(x_4,p_{2t})F_{2g}(\Delta,x_3)F_{2g}(\Delta,x_4)
 \eeq
 where the two gluon form factor
 \beq
 F_{2g}(\Delta)=\frac{1}{(1+\Delta^2/m^2_g)^2}
 \eeq
 and the parameter
 \beq
 m_{g}^2=8/\delta ,
 \eeq
 where
 \beq
 \delta=max(0.28fm^2, 0.31fm^2+0.014fm^2\log(0.1/x)),
 \eeq
 and was determined from the analysis of the exclusive $J/\Psi$ diffractive photoproduction\cite{Frankfurt}.
 The functions $G$ are the gluon pdf of the proton, which we parameterize using
\cite{GRV}.
 Then
 \beq \int \frac{d^2\Delta}{(2\pi)^2}U(\Delta)=\frac{1}{4\pi}\frac{m^4_g(x_3)m^4_g(x_4)}{(m_g^2(x_3)-m_g^2(x_4))^2}(1/m^2_g(x_3)-1/m^2_g(x_4)+\frac{2\log(m^2_g(x_3)/m^2_g(x_4))}{(m^2_g(x_3)-m^2_g(x_4)}
.
 \eeq
 In the limit $x_3\sim x_4$ we recover
 \beq \int \frac{d^2\Delta}{(2\pi)^2}U(\Delta)=\frac{m^2_g}{(12\pi)}.
 \eeq
 \subsubsection{The $\gamma A$ case.}
 The general expressions for a nuclear target is
\beq
\frac{d\sigma}{dx_1x_2dx_3dx_4dp_{1t}^2dp_{2t}^2}=D(x_1,x_2,p_{1t}^2,p_{2t}^2)G(p_{1t}^2,x_3)G(p_{2t}^2,x_4)\frac{d\sigma}{dt_1}\frac{d\sigma}{dt_2}\int d^2\Delta F'_A(\Delta,-\Delta )
\eeq
 where
\beq
F'_A(\Delta,-\Delta )=F_A(\Delta,-\Delta )+AU(\Delta).
\label{tup}
\eeq
Here $F_A(\Delta,-\Delta )$  is the  nucleus body form factor, and the form factor $U$ was defined in  Eq. \ref{U}. The first term in Eq. \ref{tup}
corresponds to the processes when two gluons
originate
 from the different nucleons in the nucleus
 while the  second term in Eq.\ref{tup} corresponds to the case when they originate
 from the same nucleon. The first term is expected to dominate for heavy nuclei as it scales as
  $A^{4/3}$ \cite{ST,BSW}.
 \par For the nuclear target we have
 \beq
 F_A(\Delta,-\Delta)=F^2(\Delta), F(\Delta)=\int d^2b \exp(i{\vec \Delta}\cdot  {\vec b})T(b),
 \eeq
 and
 \beq
 T(b)=\int dz \rho_A(b,z)dz
 \eeq
 is the nucleus profile function.
 The nuclear form factor integral is expressed through the  profile function as
  \beq \int \frac{d^2\Delta}{(2\pi)^2}F(\Delta,-\Delta)=\int T^2(b)d^2b=\pi\int T^2(b)db^2,
  \eeq
  where T(b) is calculated using the conventional mean field nuclear density \cite{Bohr}
 \beq
\rho_A(b,z)=\frac{C(A)}{A}\frac{1}{1 + \exp{(\sqrt{b^2 + z^2} - 1.1\cdot A^{1/3})/(0.56)}}.
\eeq
The factor $C(A)$ is a normalization constant
\beq
\int d^2bdz \rho_A(b,z)=1.
\eeq
 Here the  distance scales  are given in fm.

 There can be also the ladder splitting from the proton side.
 However
  such process corresponds to 2 to 4 process in notations of \cite{BDFS1}
 and thus does not contribute to MPI in the LLA we consider here. Such processes 
 constitute
 $\alpha_s$ corrections to conventional 2 to 4
 four jet production, and it is expected they give a small contribution in the back to back kinematics.
 This is consistent with the  results of modeling a tree level processes 2 to 4 in $p\bar p$  in Tevatron carried out by D0 and CMS and Atlas at LHC-see ref. \cite{Tevatron1,Tevatron2,Tevatron3,cms1,cms2,cms3,Atlas}.
 Still this issue definitely deserves a further study. The relative rate of MPI and 2 to 4 processes plays an important role in accessing   feasibility of  observing  MPI.
\subsubsection{The ratio of MPI events to dijet rate for $pA$ and $AA$.}
\par The
$pA$ collisions are dominated  by the
 $Ap$ process where  a much larger   flux factor is generated
 by projectile nuclei leading to dominance of the ultra peripheral collisions of photons with protons.  Hence in such  process one predominantly measures   a double GPD of a proton.
At the same time in the ultraperipheral $AA$ process the dominant contribution originates from the  interaction of charmed pair with two gluons coming from different protons \cite{ST,BSW}.
The ratio of cross section of such DPI process  in $AA$ scattering  to the
cross section of DPI cross section in  $pA$ scattering, in which  both gluons belong to the same  nucleon is (since the photon  flux from nuclei is the same)
\beq
\frac{\frac{Am^2_g}{12\pi}}{\int F_A(\Delta,-\Delta)\frac{d^2\Delta}{(2\pi)^2}}\sim 2,\label{simon1}
\eeq
where we take 
 $A=200$. Thus the ratio of the  total  number of the MPI events in  $AA$ to the rate calculated in the impulse approximation is $\sim 3$.This is consistent with a numerical analysis that shows that for the same c.m. energies
the ratio of number of MPI events to dijet rate in $AA$ collisions is 2.5--3 times larger than the same ratio for $pA+Ap$ process.

\par Note that this result is purely geometrical, we find a similar 
ratio \ref{simon1} 
for $\gamma p$ and $\gamma A$ collisions at the  LHeC.
\subsection{Hard matrix elements.}
The cross sections  $d\sigma/dt$  are usual dijet cross sections calculated with $s\rightarrow s_{\gamma N}=2k\sqrt{s}$, where $s$ is the invariant energy of the $ep(AA,pA)$.
.
We have
\beq
d\sigma/dt=\frac{(4\pi\alpha_s(Q^2))^2M^2}{(x_1x_3\sqrt{x_1x_3})16\pi s^{3/2}\sqrt{x_1x_3s-4p_t^2}},
\eeq
where $Q^2$ is the dijet transverse scale.
Here the matrix element $M$ of the $c$-quark - gluon scattering is given by
\beq
M^2=(4\pi\alpha_s(Q^2))^2(-\frac{4}{9}(\frac{\hat u}{\hat s}+\frac{\hat s}{\hat u})+\frac{\hat t^2+\hat u^2}{\hat s^2}),
\eeq
\noindent
where
 \beq \hat s =x_1x_3s, \hat t=-s(1-z), \hat u=-\hat s (1+z), z=\cos\theta=\tanh(y_1-y_3)/2=\sqrt{1-4p_{1t}^2/(x_1x_3s)}.
\eeq
The angle $\theta$ is the scattering angle in the c.m. frame of the dijet.
The region of integration is given by
\beq
x_1x_3s-4p_t^2>0,x_1>0.2
\eeq
The integration over the second dijet event goes in the same way, with $x_1\rightarrow x_2,x_3\rightarrow x_4$.
\par For two dijet event, in order to find the event rate,  the cross section calculated above must be convoluted with photon flux determined using Weiczsacker - Williams approximation:

\beq
N_2=\int dp_{1t}^2dp_{2t}^2 du\frac{dN}{du}\int \frac{d\sigma}{dx_1dx_2dx_3dx_4}dx_1dx_2dx_3dx_4,
\eeq
where the limits of integration are determined by $x_1x_3s-4p_{1t}^2>0$, $x_2x_4s-4p_{2t}^2>0$, $x_1,x_2>0.2$.
In the same way we calculate the rate for production of one pair of jets:
\beq
N_1=\int dp_{1t}^2 du\frac{dN}{du}\int \frac{d\sigma}{(dx_1dx_3)}dx_1dx_3.
\eeq
The limits of integration are determined by $x_1x_3s-4p_{1t}^2>0$.

\subsection{LHeC/HERA kinematics - $ep$ collisions.}
For $ep$ collisions we use the standard variable $y$
\beq s_{\gamma p}/s=y
\eeq
The  photon flux is
\beq
dN/dy=\frac{\alpha_{em}}{2\pi}(\frac{1+(1-y)^2}{y})\log(Q^2_{max}/Q^2_{min})-2m^2_ey(1/Q^2_{min}-1/Q^2_{max}),
\eeq
where $Q^2_{max}\sim 1$ GeV$^2$, and $Q^2_{min}=4m^2_ey/(1-y)$.

\subsection{Ultraperipheral collisions at the LHC: $AA$ and $pA$ cases.}
\par

In ultraperipheral collisions two nuclei (proton and nucleus) scatter at large impact parameters with one of the colliding particles  emitting a Weizsacker Williams  photon which interacts with the second particle producing two jets in the $\gamma+p_2 \to 4$ jets$ +X$ reaction
( for a detailed review see \cite{Baltz:2007kq}).
The corresponding total cross section is
calculated by convoluting the elementary $\gamma N$ cross section  with the flux factor
\beq
\frac{dN}{dk}=\frac{2Z^2\alpha_e}{\pi k}(wK_0(w)K_1(w)-\frac{w^2}{2}(K_1^2(w)-K_0^2(w))),\label{tup21}
\eeq
where $w=2kR_A/\gamma_L$, $\gamma_L=\sqrt{s_{NN}}/(2m_p)$,$s_{\gamma N}=2k\sqrt{s_{NN}}$. For the proton-nucleus reactions the flux
is described by the same Eq. \ref{tup21}, the only difference is that in definition of $w$ we substitute $2R_A\rightarrow R_A+r_p$
  where $r_p$ is the proton radius.
 In the second process  the dominant contribution is the interaction of the   a photon radiated by a heavy nucleus with the  proton. The factor in the square brackets accounts for the full absorption at impact parameters
$b<2R_A$( $b<r_p+R_A$ for $pA$ scattering).
The Bjorken fractions in the previous section where calculated relative to $s_{\gamma p}$. In order to calculate
the
total inclusive
cross section we must integrate over $k$ from $k_{min}$ corresponding to minimal $k$ necessary   to produce  four jets in the discussed kinematics
up to  $k_{max}$, $2k_{max}\sqrt{s_{NN}}=2E_mm_p$, and $E_m=\gamma_L/R_A$. We must fix $x_1$. Then we have to calculate the cross section at $x_1'=x_1/z$
where $z=k/s_{NN}$. Thus  to determine the total inclusive
  cross section for given $x_1$, we have to integrate over $k$,
substituting in  the formulae of the previous section,  $x_1',x_2'=x_{1,2}\sqrt{s_{NN}}/k $  instead of $x_1,x_2$.
The integration region is $x_1\sqrt{s_{NN}}<k<k_{max}$.

\section{Numerics.}
\par Since there is  no corresponding 4 to  4 process, it makes no sense to define $\sigma_{eff}$ for these collisions as it is usually done in the studies of MPI in $pp$ scattering.
Instead we will  calculate the number of MPI events as a function of jet
cutoff - starting from 5 GeV, as well as  the ratio
of MPI events to a total number of  dijet events with the same cutoffs for $ep$, $Ap$ and $AA$ collisions.
\par In all cases we observe a rather rapid decrease of MPI rate as a function of $p_t$. In order to calculate the rates we use
the GRV structure functions for proton \cite{GRV} and the GRV structure function for photon \cite{reya1,reya2,reya3}.
\subsection{Direct photon MPI at HERA and LHeC.}
\begin{center}
\begin{figurehere}
 \includegraphics[height=5.5cm]{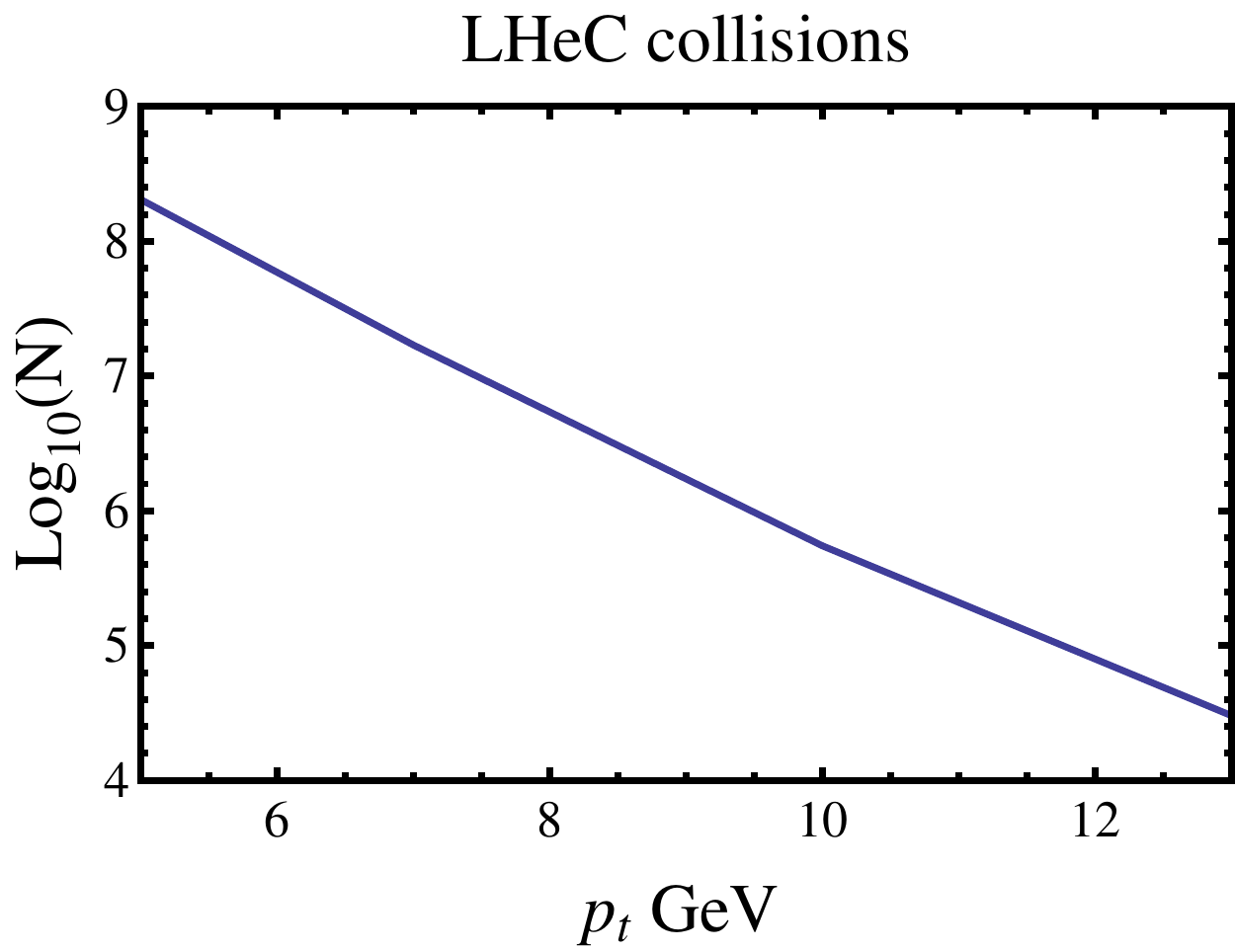}
\caption{\label{Fig5} Event rate for $ep$ MPI collisions as a function of $p_t$ cut at LHeC.}
\end{figurehere}
\end{center}
  \begin{center}
 \begin{figurehere}
   \includegraphics[height=5.6cm]{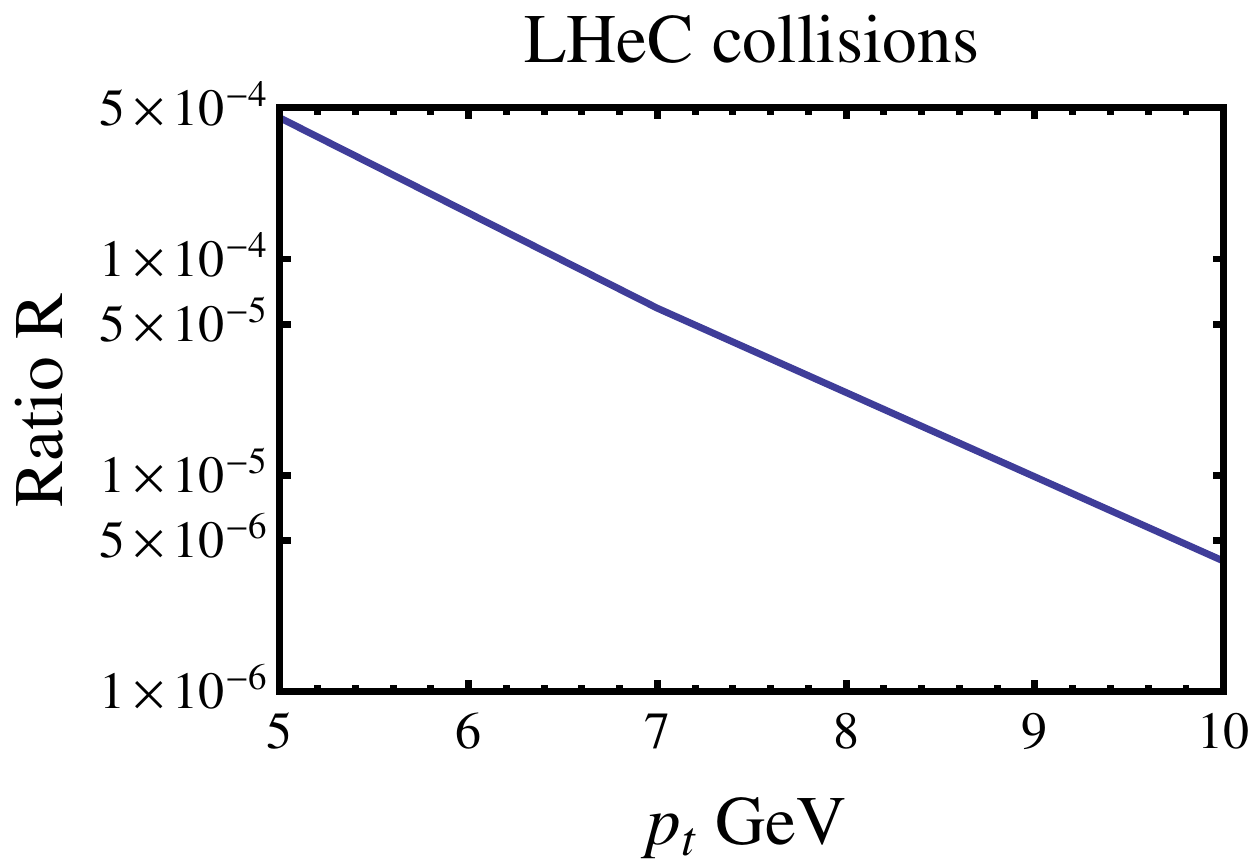}
   \caption{  \label{kin1}The ratio of a number of MPI events to the of dijet rate, $\sqrt{s}=1.3$ TeV.}
 \end{figurehere}
 \end{center}
 To estimate the MPI event rate at LHeC at $\sqrt{s}=1.3$ TeV we used luminosity $10^{34}$ $cm^{-2}s^{-1}$.
 The number of events and their ratio to the total number of dijet event are 
 presented in Figs. 5,6.
For cutoff $p_t>5$ GeV we get $2\cdot 10^8$ events for 
the running time of $10^6$ s.
 The ratio to a number of dijet events with the same cutoffs on $x$ and $p_t$ is $0.045\%$.

We also considered MPI event rate in the  similar kinematics at HERA.
To estimate the MPI event rate at HERA we use the total integrated luminosity accumulated at HERA of $1$ fb$^{-1}$,
at the  $\sqrt{s}$=300 GeV.
For cut off $p_t>5$ GeV we get $1.2\cdot 10^5$ events.  The ratio to the number of dijet events with the same cutoffs on $x$ and $p_t$ is $0.0125\%$. However,
 it  seems difficult to connect these   results with the available data on multiparton processes studied at  HERA \cite{Jung}.
 The reason is a rather  low
 efficiency of selecting events with  charm  production. For example, the total
 number of events with charm identified in the ZEUS study was   11000 $D^*$ events corresponding to a handful of MPI events of the type we discuss \cite{Zeus}.

\subsection{AA collisions.}
\begin{center}
\begin{figurehere}
 \includegraphics[height=5.5cm]{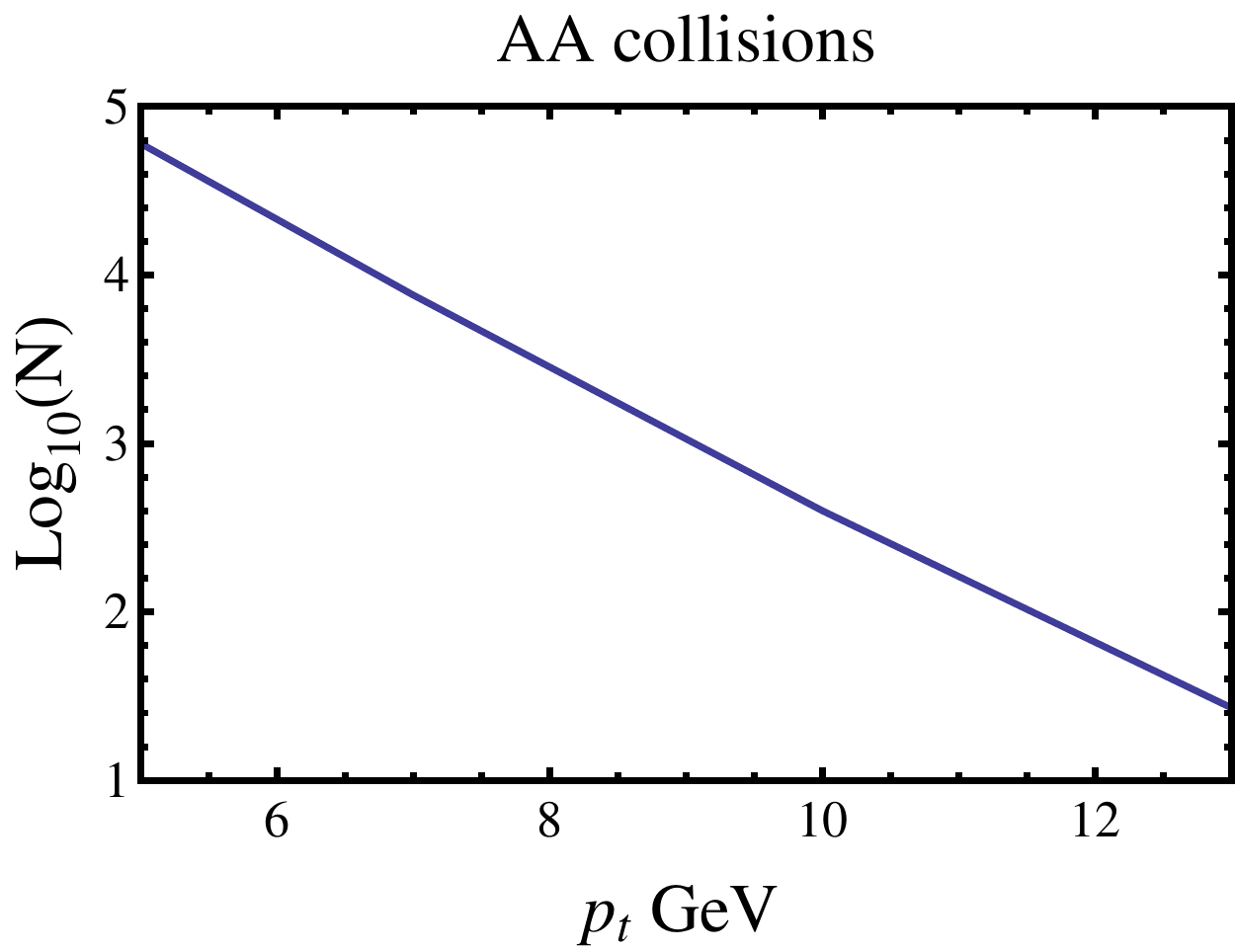}
\caption{\label{Fig20}The event rate for MPI in $AA$ collisions. }
\end{figurehere}
\end{center}

\begin{center}
\begin{figurehere}
 \includegraphics[height=5.5cm]{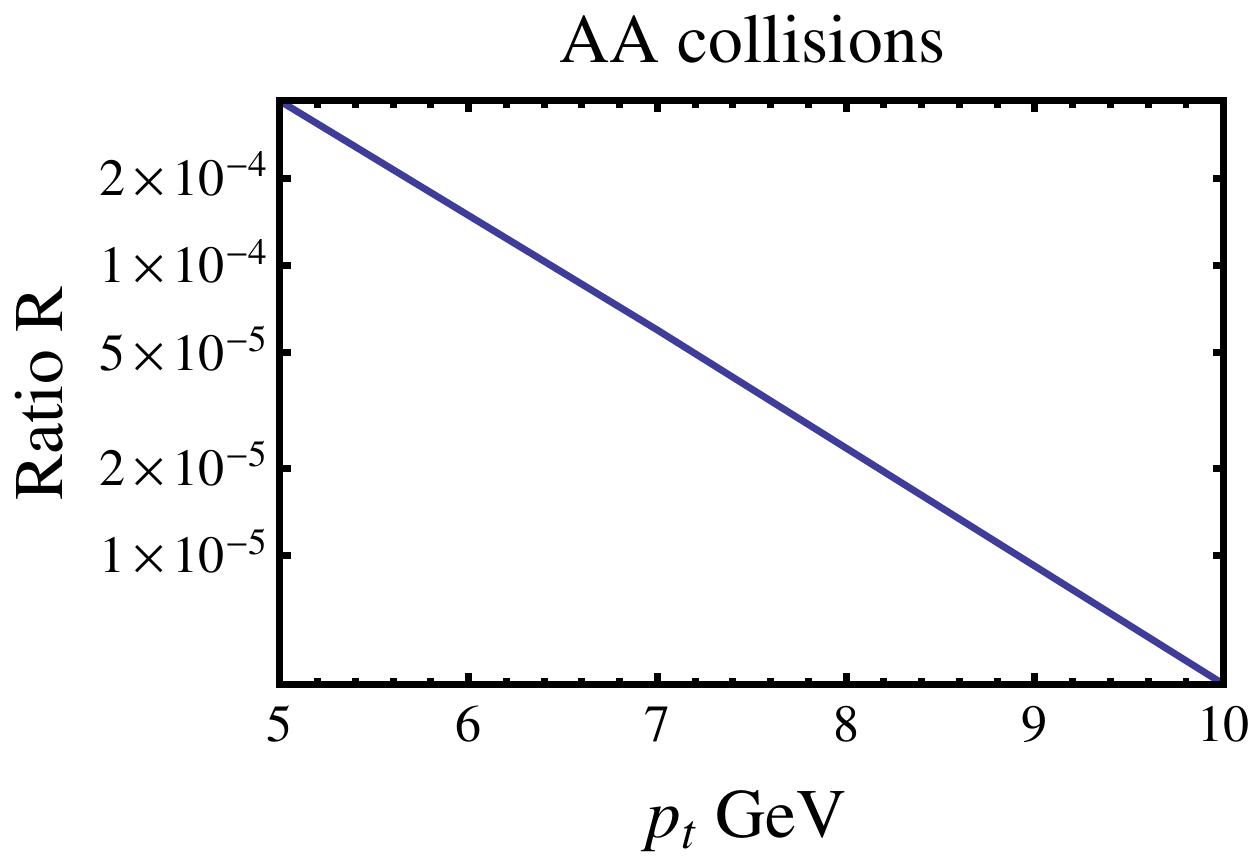}
\caption{\label{Fig30}The ratio of MPI and the  dijet event rate as a function of minimal $p_t$ for $AA$ collisions.}
\end{figurehere}
\end{center}

For $AA$ collisions we use: (i) luminosity $10^{27}cm^{-2}s^{-1}$, (ii) running time $10^6$ s, and $\sqrt{s}=5.6$ TeV, $\gamma= E_p/m_p= 2.8\cdot 10^3$ The radius of the lead nucleus is 6.5~fm. The exponentially decreasing Macdonald function cuts off the contribution of high photon energy. The total number of the events for the $p_t$ cut of  $5$ GeV is $5\cdot 10^4$, while
the ratio of MPI events to the total number of dijet events is relatively  high--$0.037\%$
 (cf. discussion in sec. III.3).
  The number of the MPI events decreases as a function of cutoff slightly faster than
as $1/p_t^8$ , while the ratio of MPI to total number of dijet events  scales approximately as $1/p_t^4$. The number of events
and the ratio of a number of MPI and dijet events are presented in Figs. 7,8.
\par Recall that $x_{\gamma}=x_1+x_2 $ in Figs. 7,8 is  0.4.  The dependence of $\log(N)$ on $x_{\gamma}$ is shown in Fig. \ref{figx} . We see that for large 
$x_{\gamma}$
the number of events rapidly
decreases,
 while for $x<0.4-0.5$ the 
 decrease
 becomes much less rapid. If we increase $x_{\gamma}$ to 0.8 the number of $AA$ MPI events decreases by a factor of four only.
  So it
 may be possible  to focus on
 the  higher $x_\gamma$ regions where contribution of resolved photon is very small, while loosing relatively small fraction of events.
\begin{center}
\begin{figurehere}
 \includegraphics[height=5.5cm]{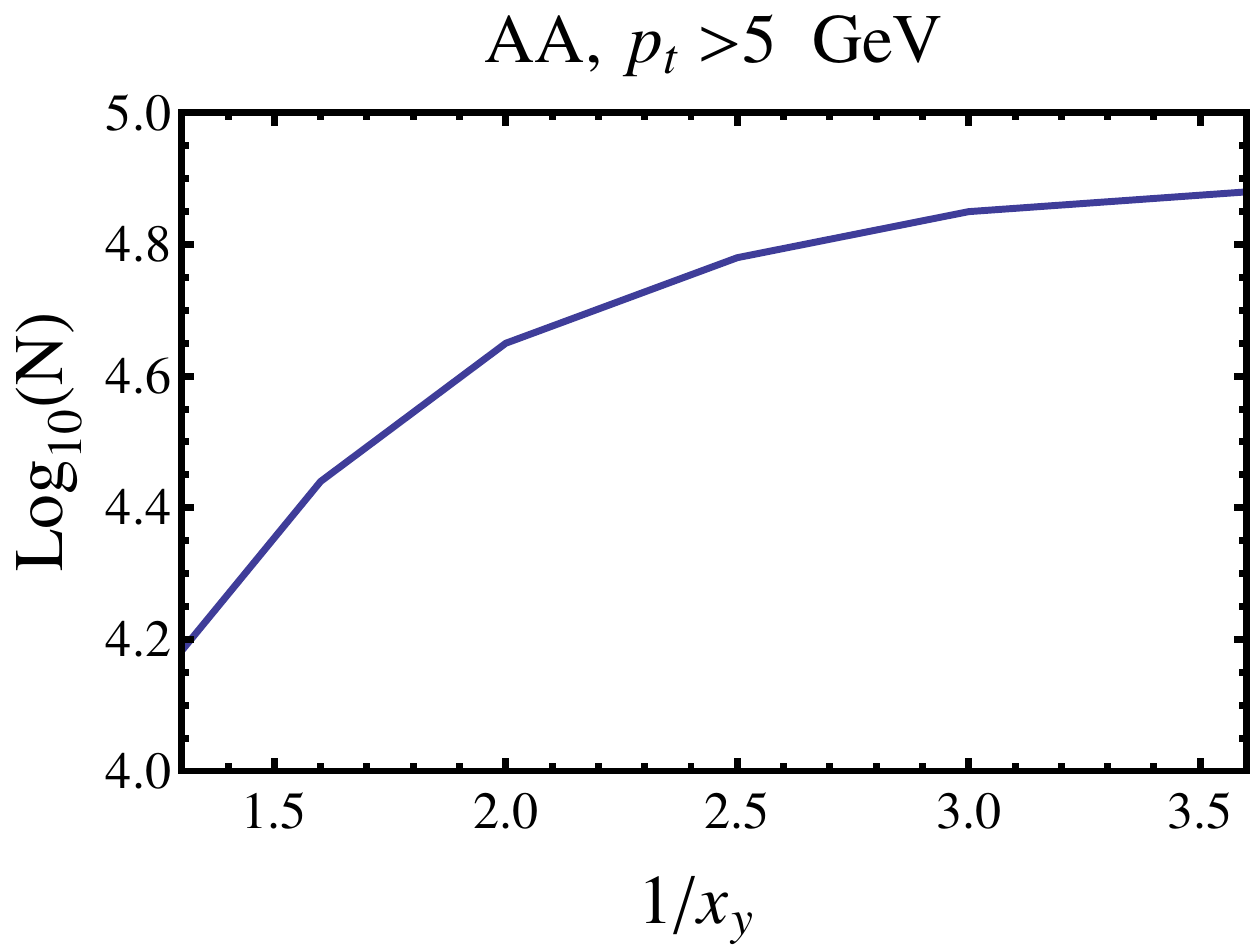}
\caption{\label{figx}The number of events as a function of cut off in $1/x_\gamma$.}
\end{figurehere}
\end{center}
Finally note that this ratio rapidly increases with energy.  If we for example take  $AA$ energies equal to  those of $pA$ scattering (i.e. a factor of 2.5
increase of $s$) the ratio increases by $30\%$.
\subsection{$pA$ collisions.}
\begin{center}
\begin{figurehere}
 \includegraphics[height=5.5cm]{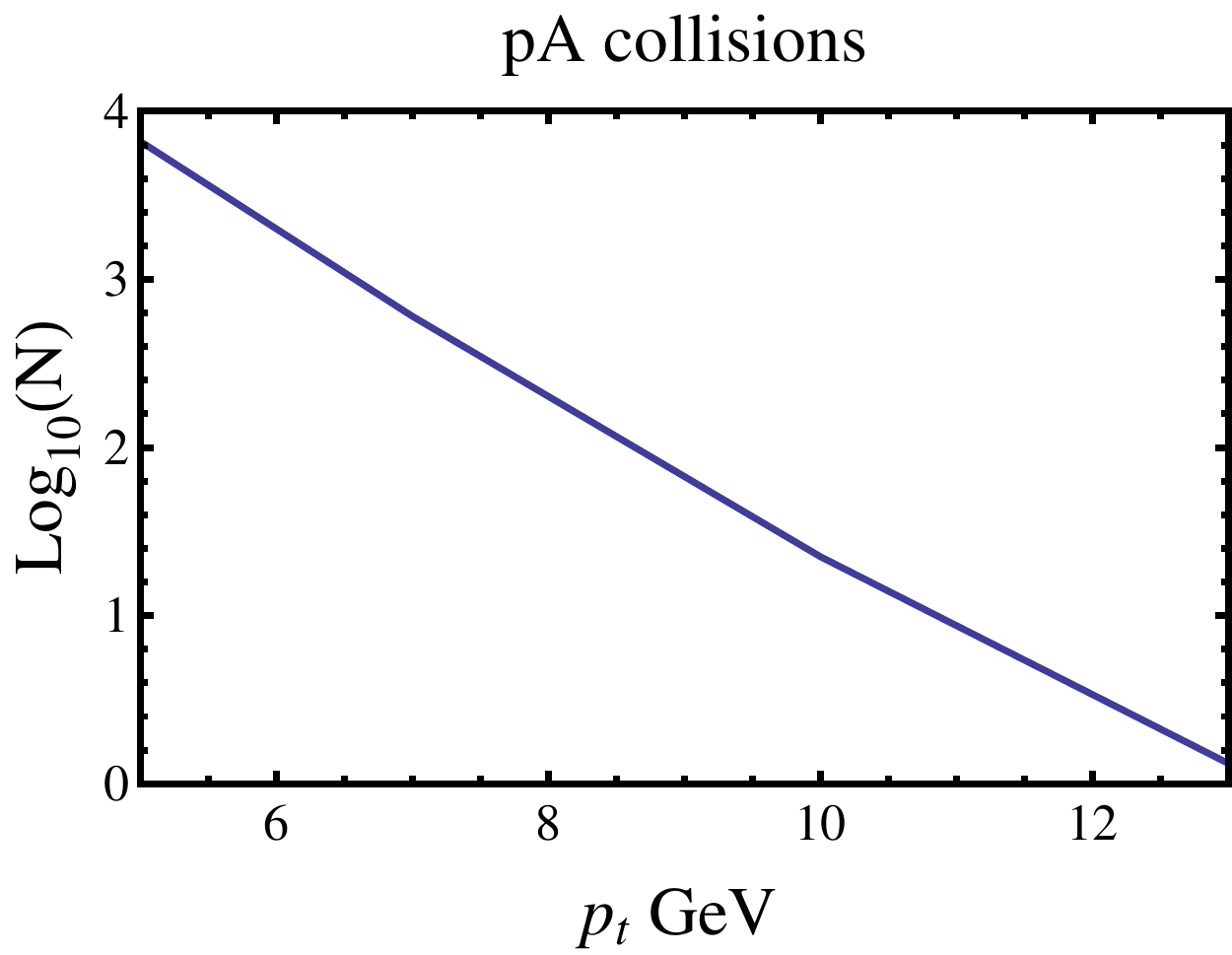}
\caption{\label{Fig22}The event rate for MPI in $pA$ collisions }
\end{figurehere}
\end{center}

\begin{center}
\begin{figurehere}
 \includegraphics[height=5.5cm]{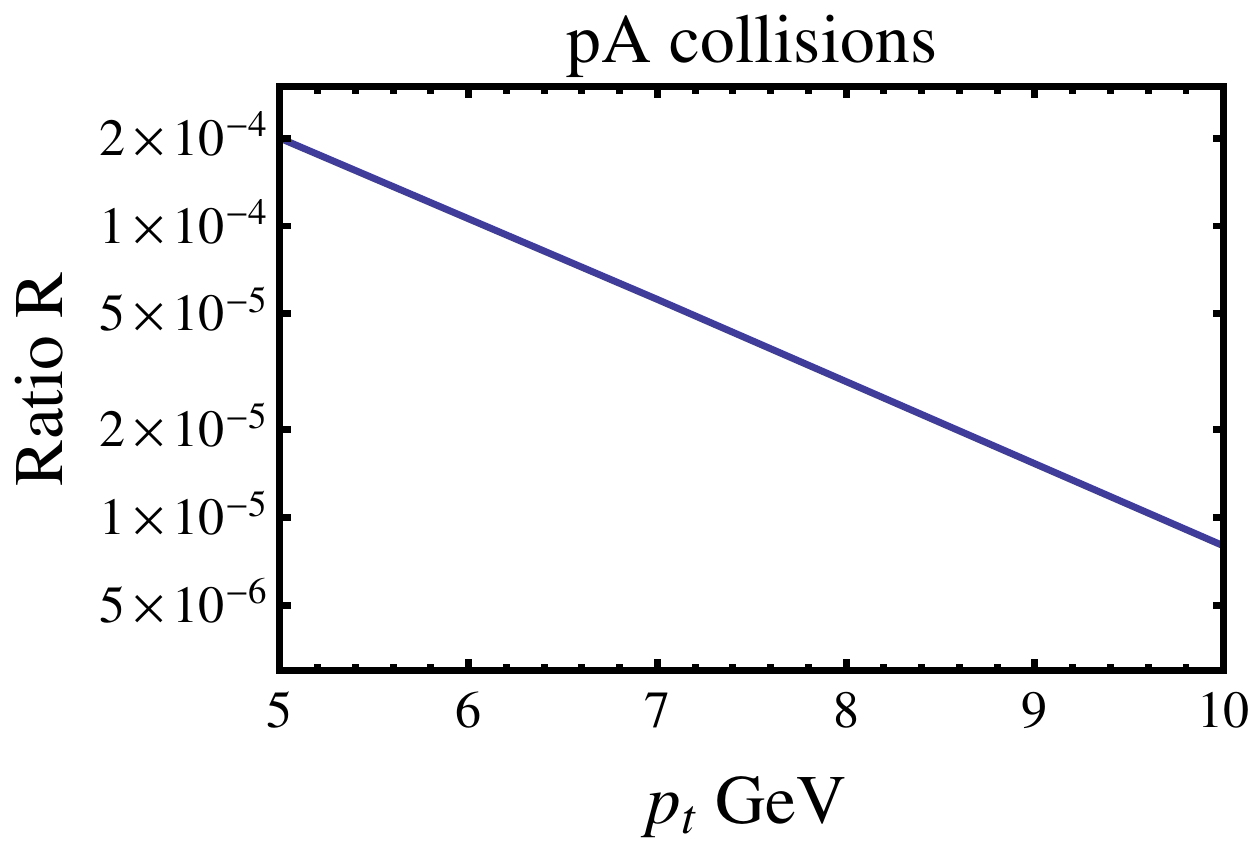}
\caption{\label{Fig32}The ratio of MPI and  dijet rate for $pA$ collisions.}
\end{figurehere}
\end{center}
For $pA$ collisions we use luminosity  $10^{29}cm^{-2}s^{-1}$, running  time $10^6$ s, and   $\sqrt{s}=9 $ TeV.
 The number of events for cut
$5$ GeV is of the order $6.6\cdot 10^3$, and ratio is of order $0.02\%$, rapidly decreasing as in the $AA$ case with increase of $p_t$. The corresponding numbers 
are shown  in Figs. 10,11 as a function of the  jet cutoff.

 From the comparison of Figs. 9 and 11 one can see that there is a factor of $\sim 2$  enhancement of the  ratio
 DPI to dijet events in  $AA$ scattering, relative to $pA$ case which is due to a combination of the geometrical enhancements we discussed above and
 suppression due to the smaller energy per nucleon in $AA$ case.

\section{Conclusions.}
\par We  derived general equations for MPI processes with production of charm  in direct photon hadron (nucleon,nuclei) collisions and used them to calculate the corresponding rates and compare them
with dijet rates. We demonstrated that the discussed processes directly measure  nucleon and nucleus $_2GPD$s.
We found a significant enhancement of the MPI / dijet cross section ration  in the $\gamma A $ scattering as compared to $\gamma p$ scattering due to the scattering off two nucleons along the photon impact parameter.
\par The analysis was done  for jet photoproduction in  the realistic kinematics, of production of two pairs  charm-gluon dijets with $p_t>5 $ GeV, and cut of $x_1,x_2>0.2$, ensuring they are created mainly
due to the direct photon mechanism. We considered these
 MPI processes for $ep$ collisions at the LHeC and HERA and for  $AA$ and $pA$ collisions at LHC and $ep$ collisions at HERA.  We conclude that the studies would definitely be feasible at the LHeC. In the case of the LHC the rates appear reasonable, and the key question is the efficiency  of the LHC detectors.
Further studies of the feasibility of the measuring discussed processes at the LHC is highly desirable. Here we just notice that
since a larger fraction of charm in the discussed processes is produced at the central rapidities  we expect  the efficiency of the detection of the discussed process would be pretty high for ATLAS and CMS.

\section*{Acknowledgements}
 We thank the CERN  theory division for hospitality during the time the initial part of this work was done.  M.S.'s
 research was supported by the US Department of Energy Office of
Science, Office of Nuclear Physics under Award No.
 DE-FG02-93ER40771.

\end{document}